\documentclass[numberedappendix]{emulateapj}

\shorttitle{Ram Pressure Stripping} \shortauthors{Hester}
\journalinfo{Hester, J. A. 2006, ApJ, 647, 910}

\begin{document}
\title{Ram Pressure Stripping in Clusters and Groups}

\author{J.A. Hester\altaffilmark{1}}
\email{jhester@princeton.edu}

\altaffiltext{1}{Jadwin Hall, Princeton University, Princeton, NJ,
08540}

\begin{abstract}
Ram pressure stripping is an important process in the evolution of
both dwarf galaxies and large spirals. Large spirals are severely
stripped in rich clusters and may be mildly stripped in groups.
Dwarf galaxies can be severely stripped in both clusters and
groups.  A model is developed that describes the stripping of a
satellite galaxy's outer H \textsc{i} disk and hot galactic halo.
The model can be applied to a wide range of environments and
satellite galaxy masses. Whether ram pressure stripping of the
outer disk or hot galactic halo occurs is found to depend
primarily on the ratio of the satellite galaxy mass to the mass of
the host group or cluster. How the effectiveness of ram pressure
stripping depends on the density of the inter-group gas, the dark
matter halo concentrations, and the scale lengths and masses of
the satellite components is explored.  The predictions of the
model are shown to be well matched to H \textsc{i} observations of
spirals in a sample of nearby clusters. The model is used to
predict the range of H \textsc{i} gas fractions a satellite of
mass $M_{v,sat}$ can lose orbiting in a cluster of mass
$M_{v,gr}$.
\end{abstract}


\section{INTRODUCTION}
Ram pressure stripping was first proposed by \citet{Gunn Gott} to
explain the observed absence of gas-rich galaxies in clusters.
They noted that galaxies falling into clusters feel an
intracluster medium (ICM) wind. If this wind can overcome the
gravitational attraction between the stellar and gas disks, then
the gas disk will be blown away.  They introduced a simple
analytical condition to determine when gas is lost:
\begin{equation}
\rho_{ICM}v_{sat}^{2} < 2 \pi G\sigma_{\ast}\sigma_{gas}.
\end{equation}
The right-hand side, $2 \pi G\sigma_{\ast}\sigma_{gas}$, is a
gravitational restoring pressure, $\rho_{ICM}$ is the density of
the ICM, and $v_{sat}$ is the orbital speed of the satellite
galaxy. Using this condition they concluded that spirals should
lose their gas disks when they pass through the centers of
clusters.

Galaxies near cluster centers have earlier type morphologies, are
redder, and form fewer stars than galaxies in the field of similar
luminosity~\citep{Gomez et al, Goto et al, Hogg et. al., Hogg et.
al. 2}.  Stripped galaxies have low H \textsc{i} masses for their
size and morphology.  Such H \textsc{i} deficiency correlates with
lower star formation rates (SFRs)~\citep{Gavazzi}, and lowered
SFRs lead to redder colors. Therefore, ram pressure stripping, in
conjunction with processes like harassment that can alter the
structure of a galaxy, may play a role in transforming galaxy
morphologies.

Several sets of observations indicate that stripping occurs to
spirals in clusters.  \citet{Giovanelli and Haynes 1983} define a
deficiency parameter that compares a galaxy's observed H
\textsc{i} mass to the expected H \textsc{i} mass in a field
galaxy with the same morphology and optical size.  Galaxies in
Virgo,~\citep{Giovanelli and Haynes 1983},
Coma,~\citep{Bravo-Alfaro et al 2000}, and other nearby
clusters,~\citep{Solanes et al 2001} are observed to be H
\textsc{i} deficient. \citet{Solanes et al 2001} observe that the
average deficiency increases with decreasing distance to the
clusters' centers and that deficient galaxies have, on average,
more eccentric orbits.  \citet{Cayatte et. al.} observe
indications of stripping in the H \textsc{i} distributions of
individual galaxies in the Virgo cluster.  Compared to an average
profile for a field galaxy of similar morphology and optical
extent, these galaxies have normal gas densities in their inner
disks but are sharply truncated.  More detailed observations of
several Virgo spirals have since been carried out.  NGC 4522 is
observed to have an undisturbed optical disk, a severely truncated
H \textsc{i} disk, and extra planar H$\alpha$ and H \textsc{i} on
one side of the disk~\citep{Kenney and Koopmann, Kenney van Gorkom
Vollmer 2004}.  IC 3392, NGC 4402, NGC 4419, and NGC 4388 all have
a truncated H \textsc{i} disk, and the first three have
extraplanar H \textsc{i}~\citep{Kenney et al 2004}.  NGC 4569 has
a truncated H \textsc{i} disk and extra planar H$\alpha$
~\citep{Kenney et al 2004}.  Observing asymmetric, extraplanar gas
in combination with an undisturbed optical disk is a strong
indication that the galaxy's gas is interacting with the ICM.

Detailed simulations of ram pressure stripping confirm that
spirals in clusters should be stripped.  For some simulations the
final radii of the galaxies are compared to the predictions of the
Gunn and Gott condition.  In addition, the morphologies of
simulated galaxies have been compared to observations of spirals
in Virgo. Simulations by~\citet{Abadi Moore Bower} show that
galaxies in a face-on wind are stripped to radii near those
predicted by the Gunn and Gott condition, but galaxies in an
edge-on wind are only mildly stripped.  In contrast, other
simulations show a two-step stripping process that strips galaxies
at all inclination angles. Ram pressure stripping first quickly
strips gas from the outer disk, and then the remaining disk is
slowly viscously stripped~\citep{Marcolini Brighenti D'Ercole,
Quilis Moore Bower, Roediger and Hensler, Schulz and Struck 2001}.
\citet{Marcolini Brighenti D'Ercole} point out that, unlike ram
pressure stripping, viscous stripping is more effective in an
edge-on wind. They find that the final radii of the face-on
galaxies match those predicted using the Gunn and Gott condition
and that edge-on galaxies are stripped to slightly larger radii.
\citet{Roediger and Hensler} run a series of simulations to test
the dependence of the stripping radius on the speed of the ICM
wind, the mach number, and the vertical structure of the gas disk.
They find that the most important parameters are the ram pressure
and surface density of the gas disk and that the effects of
varying the mach number and vertical structure are minor.  They
find that the original Gunn and Gott condition does a fair job
predicting the stripping radius. However, when they use an
adjusted ram pressure, they find that using the thermodynamic
pressure in the central plane of the disk more accurately predicts
the stripping radius.

Simulations of ram pressure stripping can also match the range of
morphologies of stripped galaxies.  \citet{Vollmer et al 2000,
Vollmer et al 2004} track the density, velocity, and velocity
dispersion of a simulated galaxy's cold disk gas.  In NGC 4522,
the kinetically continuous extraplanar gas matches simulations of
a galaxy that is currently being stripped.  However, the gas in
NGC 4569 is less pronounced, not kinetically continuous with the
disk gas, and has a large velocity dispersion, more like
simulations of a galaxy that has been stripped in the past. In
their simulations~\citet{Schulz and Struck 2001} observe a process
they term "annealing".  The ICM wind compresses the gas disk and
triggers the formation of large spiral arms. The interaction of
these arms with the wind leads to the contraction of the galaxy
and the formation of a dense truncated gas disk.  A burst of star
formation may occur at the edge of the truncated disk.  A ring of
star formation is seen at the truncation radius of the disk in IC
3392.

Some groups have studied stripping in groups and cluster outskirts
where the ram pressure is lower than in cluster centers. While
simulations and observations both support the idea that large
spirals in clusters are stripped of their H \textsc{i} disks, it
is not clear that stripping occurs in poorer environments.  Two
types of stripping may occur in groups.  The ram pressure in
groups is lower than in clusters.  However, dwarf galaxies have
lower masses than large spirals and therefore lower restoring
pressures, and dwarf galaxies may be stripped of most of their H
\textsc{i}. In addition, large spirals have restoring pressures
that decrease with radius and may be stripped of their outer
disks. \citet{Marcolini Brighenti D'Ercole} run simulations of
dwarf-like galaxies in winds typical of the ram pressure in groups
and find that the simulated dwarfs are stripped. The
\citet{Roediger and Hensler} simulations include runs with low ram
pressure.  Their large and medium-sized model spirals are stripped
of some of their outer H \textsc{i} in these runs. The
observational evidence for stripping in low-mass systems is
circumstantial.  The gas content and star formation histories of
Local Group dwarfs correlate with their distances from the
dominant spirals~\citep{Blitz Robishaw, Grebel}.

This project aims to use the clear observations of stripping in
clusters to predict when stripping occurs in groups. This is done
by examining how the ram and restoring pressures depend on the
masses of the cluster or group and the satellite galaxy.  Two ways
in which the masses of the host system and satellite galaxy affect
the ram and restoring pressures are identified.  Satellites'
orbital speeds are set by the depth of the cluster's potential,
and the restoring force on the gas disk is determined by the depth
of the satellite's potential. In addition, both cluster and galaxy
morphologies change with mass. A model is developed and used to
study both types of mass dependence.  This is done in the
following steps.

In the next section a model for the mass and gas distributions is
presented both in terms of physical parameters and in terms of
scale-free parameters.  In section 3 the dependence of ram
pressure stripping on the masses of the satellite and cluster is
presented assuming that the scale-free parameters are constant. In
section 4 the dependence of the scale-free parameters on the
masses of the satellite and cluster and the sensitivity of the
model to these changes are explored. In section 5 the results are
discussed and the model is compared to observations of stripping.
Section 6 concludes.

\section{THE MODEL}
The analytical model of ram pressure stripping developed here uses
the Gunn and Gott condition to predict the extent to which
galaxies are stripped. This condition is based on a simplified
picture of stripping that ignores the details of the
hydrodynamical processes and balances forces on large scales.
However, simulations of galaxies that are stripped by a face-on
wind result in stripping radii that match the predictions of the
Gunn and Gott condition~\citep{Abadi Moore Bower, Marcolini
Brighenti D'Ercole, Roediger and Hensler}.  In the simulations
galaxies in an edge on wind are eventually viscously stripped to
radii similar to those of face-on, ram pressure stripped galaxies.
This is discussed further in section~\ref{sec inclination}.  In
light of the general agreement between Gunn and Gott's condition
and simulations of ram pressure stripping, it is assumed that a
model based on this condition can be used to make general
statements about where and to what extent galaxies are stripped.

The model for the group or cluster has two components, a
gravitational potential in which the satellite orbits and an ICM
against which the satellite shocks.   The gravitational potential
is modeled using an NFW profile with mass $M_{v,gr}$, radius,
$r_{v,gr}$, and concentration, $c_g$.  The scaled radius $s\equiv
r/r_{v,gr}$.
\begin{equation}\label{group NFW}
\phi_{halo}=\frac{-GM_{v,gr}}{r_{v,gr}}g(c_g)\frac{\ln(1+c_{g}s)}{s}
\end{equation}
The NFW profile is discussed in more detail in appendix~\ref{NFW
profile}. The ICM is modeled using a $\beta$ profile with scale
length $r_c$, and central density, $\rho_0$.
\begin{equation}
\rho = \rho_0\left(1 + \frac{r^2}{r_c^2} \right)^{(-3/2)\beta}
\end{equation}
The $\beta$ profile can be rewritten as
\begin{equation}\label{beta}
\rho = \frac{\alpha v_{\rho} \rho_c^0}{3}\left(1 +
\frac{s^2}{(r_c/r_{v,gr})^2} \right)^{(-3/2)\beta}
\end{equation}
The parameter $\alpha$ relates $\rho_0$ to the characteristic
density of the NFW profile, and $v_{\rho}\rho_c^0$ is the average
dark matter density within $r_{v,gr}$.

\label{sec. restoring pressure}

The gravitational potential of the satellite and the surface
density of the gas disk are needed to model the restoring
pressure.  The gravitational potential combines a dark matter
halo, a stellar disk, and a stellar bulge.  The dark matter halo
is modeled as an NFW profile with mass $M_{v,sat}$ and
concentration $c_s$. The disk, of mass $M_d$, is described using a
Miyamoto and Nagai potential~\citep{Miyamoto Nagai}.
\begin{equation}\label{stellar disk}
\phi_{disk}=\frac{-GM_{d}}{\sqrt{R^{2}+\left(R_{d}+\sqrt{z^{2}+h^{2}}\right)^{2}}}
\end{equation}
This potential is easily differentiated, and with the appropriate
choice of disk scale length, $R_d$, and height, $h$, many of the
properties of the potential of an exponential disk can be
matched~\citep{Johnston Spergel Hernquist}.  This potential can be
rewritten as
\begin{equation}\label{stellar disk 2}
\phi_{disk}=\frac{-GM_{d}}{R_{d}}\frac{1}{\sqrt{\lambda_{d}^{2}S^{2}+\left(1+\sqrt{\lambda_{d}^2z_{s}^{2}+\lambda_{h}^{2}}\right)^2}}
\end{equation}
where $\lambda_{d}\equiv r_{v,sat}/R_{d}$, $\lambda_{h}\equiv
h/R_{d}$, $S\equiv R/r_{v,sat} $, and $z_{s}\equiv z/r_{v,sat} $.

The bulge, of mass $M_b$, is described using a Hernquist
potential~\citep{Hernquist}.
\begin{equation}\label{bulge}
\phi_{bulge}=\frac{-GM_{b}}{r+r_{b}}
\end{equation}
This can be written as
\begin{equation}\label{bulge 2}
\phi_{bulge}=\frac{-GM_{b}}{r_{b}}\frac{1}{\lambda_{b}s_{sat}+1}
\end{equation}
where $\lambda_{b}\equiv r_{v,sat}/r_{b}$,  $s_{sat}\equiv
r_{sat}/r_{v}$, and $r_{sat}$ is the distance from the center of
the satellite. By introducing $m_{ds}\equiv M_{d}/M_{v,sat}$ and
$m_{b}\equiv M_{b}/M_{v, sat}$, the full gravitational potential
of the satellite can be written in the form
\begin{equation}
\label{eqn. full restoring pressure}
\phi_{full}=\frac{-GM_{v,sat}}{r_{v,sat}}f\left(\vec{s_{sat}},
m_{i}, \lambda_{i}, c_{s}\right)
\end{equation}

Gas in galaxies can be found in three spatial components: an
exponential molecular gas disk with a scale length comparable to
the stellar disk, a nearly flat atomic disk that extends beyond
the stellar disk and shows a sharp cutoff, and a hot galactic
halo. Beyond the disk cutoff, the atomic disk may continue as an
ionized gas disk~\citep{Binney Merrifield}.

The H \textsc{i} is modeled as a thin flat disk with mass $M_{g}$,
surface density, $\sigma_g$, and a sharp cutoff at radius $R_g$
\begin{equation}
\sigma_{g} = \cases{\sigma_0, & $R < R_g$ \cr 0, & $R > R_g$ \cr}
\end{equation}
\begin{equation}\label{sigma not}
\sigma_{0}\equiv\frac{M_{g}}{\pi
R_{g}^{2}}=\frac{M_{v,sat}}{r_{v,sat}^2}\frac{m_{dg}\lambda_{g}^{2}}{\pi}
\end{equation}
where $m_{dg}\equiv M_{g}/M_{v,sat}$ is the fractional mass of the
gas and $\lambda_{g}\equiv r_{v,sat}/R_{g}$ is the scaled size of
the disk.

The inner gas disk is dominated by the $H_2$ disk.  The molecular
disk is significantly more difficult to strip than the H
\textsc{i} disk both because the $H_2$ disk is more compact and
because the $H_2$ is found in molecular clouds.  These clouds do
not feel the effect of stripping as strongly as the diffuse H
\textsc{i}.  How stripping of $H_2$ occurs and how the $H_2$
clouds and diffuse H \textsc{i} in the inner disk interact are not
known.  The two phases may be tied together by magnetic fields; in
which case the inner H \textsc{i} disk will remain until the ram
pressure can remove the entire inner disk.  It may also be
possible for the wind to remove the H \textsc{i} from the inner
disk while leaving the $H_2$ behind.  Because of this uncertainty,
the stripping of the inner disk will not be studied here.  The
model is only used to discuss stripping beyond 1.5 stellar scale
lengths.  If the ram pressure is not strong enough to affect the
$H_2$, the $H_2$ will contribute to the restoring pressure in the
outer disk. Therefore, the mass of the $H_2$ disk is added to the
stellar mass.

The emphasis of this project is to study the stripping of the H
\textsc{i} disk.  However, the fate of the hot galactic halo can
also influence the evolution of the stripped galaxy. It is
expected that the hot halo will be easily stripped both because it
is diffuse and because the restoring potential of the satellite is
strongest in the disk.  To check this a hot halo is included in
the model galaxy.

\citet{Mori Burkert} give an analytic condition for the complete
stripping of a hot galactic halo.
\begin{equation}\label{halo stripping condition}
  \rho_{ICM}v^2_{sat}>P_{0,th} = \frac{\rho_{0,sat}k_{B}T}{\mu m_p}
\end{equation}
where $P_{0, th}$ and $\rho_{0,sat}$ are the central thermal
pressure and density of the hot galactic halo, $k_B$ is the
Boltzmann constant, $T$ is the temperature of the galactic halo,
and $\mu m_p$ is the average molecular mass. The thermal pressure
replaces the gravitational restoring pressure in the original
stripping condition.  In their simulations, this condition does a
fair job of predicting the mass of a galaxy that can be completely
stripped.  This condition is adapted for the current model and all
gas outside the radius where $P_{ram} = P_{th}$ is assumed to be
stripped.

The hot galactic halo is modeled by assuming that the gas is at
the virial temperature, $T_v$, of the satellite and is in
hydrostatic equilibrium in the satellite's dark matter potential.
The self-gravity of the gas is ignored.  The density of the hot
halo is
\begin{equation}\label{g halo prof}
\rho(s_{sat}) =
\left(\frac{v_{\rho}\rho_0^c}{3}\right)m_{sg}j(s_{sat}, c_{s})
\end{equation}
where scaled density profile $j(s_{sat}, c_s)$ is defined in
appendix~\ref{ap satellite halo} and $m_{sg}$ is the ratio of the
mass of gas in the hot galactic halo to $M_{v,sat}$.

A dimensionless temperature, $t(c_{s})$ is also introduced in
appendix~\ref{ap satellite halo}.  It is defined by
\begin{equation}
t(c_{s})\equiv \left(\frac{k_B T_v}{\mu m_p}\right)
\left(\frac{r_{v,sat}}{G M_{v,sat}}\right)
\end{equation}
It is a function of only $c_{s}$.

\section{RESULTS}
In this section the dependence of stripping on the masses of the
large and satellite halos and on the model's parameters is
explicitly determined.

The maximum ram pressure, the ram pressure at the pericenter of
the galaxy's orbit, is given by
\begin{equation}
P_{ram,max}=\frac{GM_{v,gr}}{r_{v,gr}}\frac{b_{s}^{2}}{s_{0}^{2}}\frac{\epsilon}{\epsilon_{v}}
\frac{\alpha v_{\rho} \rho_{c}^{0}}{3}\left(1 +
\frac{s_0^2}{(r_c/r_{v, gr})^2} \right)^{(-3/2)\beta}
\end{equation}
where $s_0$ is the distance of closest approach, $\epsilon$ is the
orbital energy per unit mass,
$\epsilon_v\equiv-GM_{v,gr}/r_{v,gr}$, $b_s\equiv
l^2/(|\epsilon|r_{v,gr}^2)$, and $l$ is the orbital angular
momentum per unit mass. Both $b_{s}$ and $\epsilon/\epsilon_{v}$
are expected to be independent of the masses of the satellite and
the cluster.

At any point in the orbit, $s_{orbit}$, the ram pressure is
\begin{equation}
P_{ram}(s_{orbit})=\frac{2GM_{v,gr}}{r_{v, gr}}\frac{\alpha
v_{\rho} \rho_{c}^{0}}{3}p(s_{orbit},c_{g})
\end{equation}
where dimensionless pressure, $p(s,c_g),$ is defined.
\begin{equation}
p(s,c_{g}) \equiv
\left[g(c_g)\frac{\ln(1+c_gs)}{s}-\frac{\epsilon}{\epsilon_{v}}\right]\left(1
+ \frac{s^2}{(r_c/r_{v,gr})^2} \right)^{(-3/2)\beta}
\end{equation}
Equations for the orbital speeds are derived in appendix~\ref{ap
orbits}. Assuming that $\alpha$, $\beta$, $r_c/r_{v,gr}$, and
$c_{g}$ do not depend on the mass of the halo, the $M_{v,gr}$
dependence of both $P_{ram}(s_0)$ and $P_{ram}(s_{orbit})$ is
$P_{ram}\propto M_{v,gr}^{2/3}$.

For a potential of the form in eq.~\ref{eqn. full restoring
pressure}, the force per unit gas mass in the $z$ direction is
found as follows.
\begin{equation}
f_{z}=\frac{GM_{v,sat}}{r_{v,sat}}\frac{\partial}{\partial
z}f\left(\vec{s_{sat}}, m_{i}, \lambda_{i}, c_{s}\right)
\end{equation}
\begin{equation}
f_{z}=\frac{GM_{v,sat}}{r_{v,sat}^{2}}\frac{\partial}{\partial
z_{s}}f\left(\vec{s_{sat}}, m_{i}, \lambda_{i}, c_{s}\right)
\end{equation}
Assuming that the $m_i$ and $\lambda_i$ introduced in
section~\ref{sec. restoring pressure} and $c_{s}$ are constant
with mass, the mass dependence of the restoring force per unit gas
mass is given by $f_{z}\propto M_{v,sat}^{1/3}$.

For the gas disk, if the fractional mass of the gas, $m_{dg}$, and
the scaled size of the disk, $\lambda_g$, are both constant, then
\begin{equation}
\sigma_{g}\propto M_{v,sat}/r_{v,sat}^2\propto M_{v,sat}^{1/3}
\end{equation}
Combining the above,
\begin{equation}
P_{rest,max}=\sigma_{g}f_{z,max} \propto M_{v, sat}^{2/3}
\end{equation}

For any radius along the disk, $R_{str}$, there is a maximum
restoring pressure. If the maximum restoring pressure is greater
than the ram pressure, the satellite holds the gas at this radius.
The condition for the satellite holding its gas can be written as
\begin{equation}
\frac{P_{rest,max}}{P_{ram,max}}=\frac{M_{v,sat}^{2/3}}{M_{v,gr}^{2/3}}\frac{\left(\frac{\partial}{\partial
z_{s}}\right)_{R_{str}}f\left(\vec{s}_{sat}, m_{i}, \lambda_{i},
c_{s}\right)2m_{dg}\lambda_{g}^{2}}{\alpha p(s_{orbit},c_{g})}>1
\end{equation}
\begin{equation}\label{disk M ratio}
\frac{M_{v,sat}}{M_{v,gr}}>\left[\frac{\alpha
p(s_{orbit},c_{g})}{\left(\frac{\partial}{\partial
z_{s}}\right)_{R_{str}}f\left(\vec{s_{sat}}, m_{i}, \lambda_{i},
c_{s}\right)2m_{dg}\lambda_{g}^{2}}\right]^{3/2}
\end{equation}

The restoring pressure of the galactic hot halo is
\begin{equation}
P_{rest} = \left(\frac{GM_{v,sat}}{r_{v,sat}}\right)
\left(\frac{v_{\rho}\rho_0^c}{3}\right)t(c_s)m_{sg}j(s_{sat},c_s)
\end{equation}
The condition for retaining this gas within $s_{sat}$ is
\begin{equation}\label{halo M ratio}
\frac{M_{v,sat}}{M_{v,gr}} >
\left[\frac{t(c_s)m_{sg}j(s_{sat},c_s)}{2\alpha
p(s_{orbit},c_g)}\right]^{3/2}
\end{equation}

If the model's parameters: $c_{g}$, $c_{s}$,
$\epsilon/\epsilon_v$, $b_s$, $\beta$, $r_c/r_{v,gr}$, $\alpha$,
the $m_i$, and the $\lambda_i$; are all independent of the mass of
the satellite and cluster, then the fraction of gas that is
stripped from a satellite depends on the ratio
$M_{v,sat}/M_{v,gr}$.  In this case, $P_{ram}\propto
M_{v,gr}^{3/2}$ and for both the gas disk and hot galactic halo
$P_{rest}\propto M_{v,sat}^{3/2}.$

\label{model ind}

Assuming that this set of parameters is constant is equivalent to
assuming that the physical parameters scale with mass in the most
obvious way.  It assumes that component masses scale as $M_{v}$,
lengths scale as $r_v$, and the central densities of the ICM and
the satellite's hot galactic halo are constant. For any model in
which these scalings hold, the result that the extent of stripping
depends on $M_{v,sat}/M_{v,gr}$ will hold. For any cluster mass,
the scaled radius will set the ICM density and the orbital
velocity will be proportional to $M_{v,gr}/r_{v,gr}$. The
restoring pressure depends on the depth of the satellite galaxy's
gravitational potential well and the density of the gas disk.  For
a generic potential, $\phi \propto M_{v,sat}/r_{v,sat}$ and the
restoring force per unit gas mass is proportional to
$M_{v,sat}^{1/3}$. If the mass of the gas disk scales with
$M_{v,sat}$ and the disk length with $r_{v,sat}$, then the density
of the gas disk scales as $M_{v,sat}^{1/3}$ and the restoring
pressure as $M_{v,sat}^{2/3}$.

\section{MODEL PARAMETERS}
\label{parameters}
\begin{figure*}[t]\center
\epsscale{0.80}\plotone{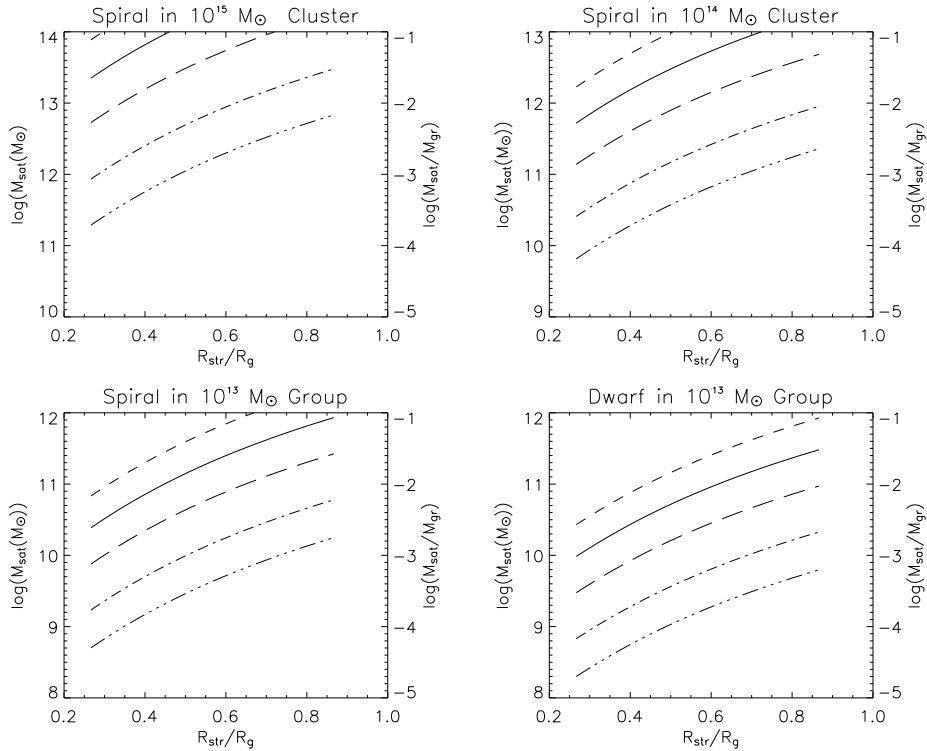} \caption{ Mass at which a
satellite's gas disk is ram pressure stripped versus
$R_{str}/R_d$. \textit{Short-dashed line}: $s_{orbit}=.25$,
\textit{solid}: $s_{orbit}=.35$, \textit{long-dash}:
$s_{orbit}=.5$, \textit{dash-dot}: $s_{orbit}=.75$,
\textit{dash-triple-dot}: $s_{orbit}=1$. \textit{Top-left}: Spiral
galaxy orbiting in a $10^{15}M_{\odot}$ cluster.
\textit{Top-right}: Spiral galaxy orbiting in a $10^{14}M_{\odot}$
cluster. \textit{Bottom-left}: Spiral galaxy orbiting in a
$10^{13}M_{\odot}$ group. \textit{Bottom-right}: Dwarf galaxy
orbiting in a $10^{13}M_{\odot}$ group.} \label{figure 1}
\end{figure*}

\begin{figure*}\center
\includegraphics[scale = 0.80]{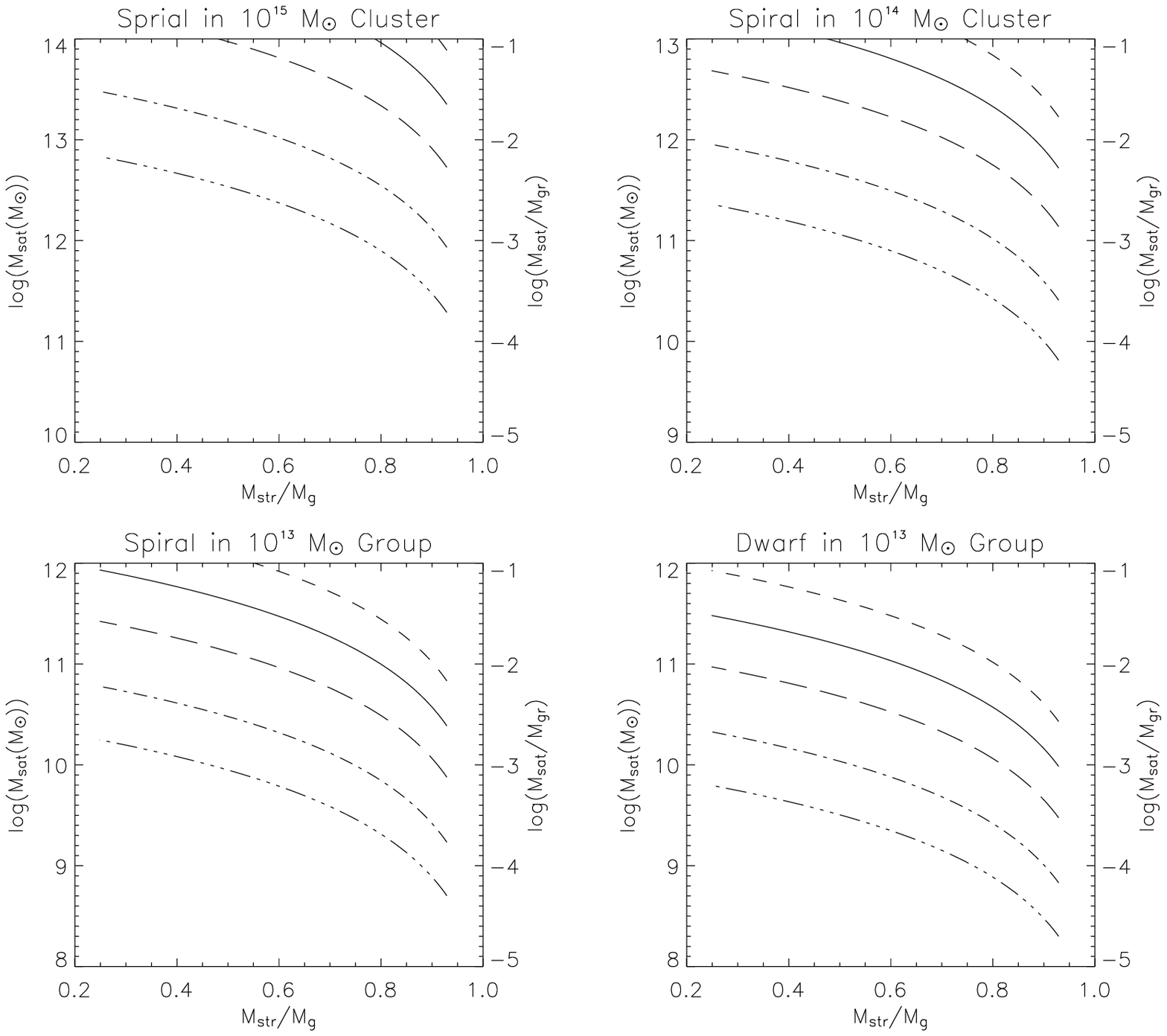}\caption{Mass at which a
satellite's disk is ram pressure stripped versus the mass fraction
of gas disk that is stripped. Lines are the same as in Fig. 1.
\textit{Top-left}: Spiral galaxy orbiting in a $10^{15}M_{\odot}$
cluster. \textit{Top right}: Spiral galaxy orbiting in a
$10^{14}M_{\odot}$ cluster. \textit{Bottom left}: Spiral galaxy
orbiting in a $10^{13}M_{\odot}$ group. \textit{Bottom right}:
Dwarf galaxy orbiting in a $10^{13}M_{\odot}$ group.}
 \label{figure 2}
\end{figure*}

In this section a reference set of values for the model's
parameters is introduced and the effect of varying these
parameters is discussed.  The reference parameters determine the
values of $M_{v,sat}/M_{v,gr}$ at which stripping occurs.  They
are taken from observations, where possible, and $\lambda$CDM
simulations, otherwise.  Two types of parameter variations are
discussed.  First, several of the model parameters vary
systematically with $M_{v,sat}$ or $M_{v,gr}$. Second, at fixed
masses the parameters have scatter.

The scatter in the parameters is discussed for several reasons. By
combining an estimate for the scatter in each parameter with the
model, the parameters that have the most effect on the radius to
which an individual galaxy is stripped can be identified. Second,
it is useful to know the range of ratios across which stripping
occurs. Finally, if the overall scatter is not too large, the
relationship between $R_{str}$ and $M_{v,sat}/M_{v,gr}$ could be
used to search for signs of stripping in a large galaxy survey.
The results of this section are discussed in the terms above in
section~\ref{predictions}.

In Figures~\ref{figure 1} and~\ref{figure 2}, the $M_{v,sat}$ at
which a satellite is stripped is plotted versus $R_{str}/R_g$ and
$M_{str}/M_{g}$ for a satellite orbiting at a variety of
$s_{orbit}$ values in $10^{15}$, $10^{14}$, and $10^{13}M_{\odot}$
clusters. The mass of the stripped gas is denoted as $M_{str}$.
Table 1 lists the model parameters used for three models, a large
satellite in a $10^{15}M_{\odot}$ cluster, the reference model; a
large satellite in a $10^{14}M_{\odot}$ cluster, the middle-mass
model; and a small satellite in a $10^{13}M_{\odot}$ group, the
low-mass model. The large satellite model is based on the Milky
Way.  Physical length scales and masses for the stellar disk and
bulge are from~\citet{Johnston Spergel Hernquist}.  The stellar
disk is modeled using $M_{ds}=10^{11}M_{\odot}$, $R_d = 6.4$ kpc,
and $h = .26$ kpc. The equivalent $R_d$ for an exponential disk is
5 kpc. The bulge is modeled using $M_b =
3.4\times10^{10}M_{\odot}$ and $r_b = .7$ kpc.  A flat disk with a
radius of 30 kpc gives $R_g/R_d$ similar to that seen in
observations.  The H \textsc{i} gas mass is chosen to result in a
disk density of $5$ $M_{\odot}$ $pc^{-2}$.  The virial mass is
chosen such that $M_{v,sat}/M_{baryon}\approx 12$. Setting
$M_{v,sat}$ also sets $r_{v,sat}$. The concentrations used are
$c_{s} = 12$ and $c_{g} = 5.5$, 7.5, and 10. The concentrations
are based on~\cite{Bullock et. al.} and assume an average
over-density within $r_v$, $v_{\rho}$, of 337. The assumed
over-density affects the concentrations, which are measured using
$v_{\rho}=337$, and $\alpha$, because it relates the ICM and dark
matter densities. Note that $v_{\rho}$ does not appear in
equation~\ref{disk M ratio} or~\ref{halo M ratio}. The ICM
parameters, $\alpha$, $\beta$, and $r_{v,gr}/r_c$, are set by
X-ray observations of clusters and groups, and the orbital
parameters are taken from simulations. Both are discussed below.
Table 1 also lists the assumed scatter in each model parameter and
resulting scatter in the mass at which a satellite is stripped.

\subsection{ICM Profile}
The profile of the ICM is determined by the parameters $\alpha$,
$r_{v,gr}/r_{c}$, and $\beta$ (eq.~\ref{beta}). \citet{Sanderson
Ponman} compile a set of average ICM profiles for clusters with
temperatures ranging from 0.3 to 17 keV.  They find that the ICM
of groups and poor clusters is both less dense and more extended
than in rich clusters. \citet{Mohr} present X-ray observations of
a set of clusters and \citet{Helsodon Ponman} and \citet{Osmond
and Ponman} present X-ray observations of groups. The three groups
present $\beta$ and $r_c$ (in units of kpc) for a best-fit $\beta$
profile and the X-ray temperature, $T_X$, for each group or
cluster.  The average $\beta$ and $r_{v,gr}/r_c$ for the cluster
observations are approximately 0.67 and 12.  The averages for the
group observations are approximately 0.45 and 170. To convert
$r_c$ (in kpc) to $r_c/r_{v,gr}$, virial masses for the groups and
clusters are found using the $M_{200}(T_X)$ relationship
from~\citet{Popesso}. The mass $M_{200}$ is the mass contained
within the radius at which the average over-density reaches 200.
It is converted to $M_v$ using equations~\ref{NFWMass} and~\ref{Mv
of rv}. The profiles in~\cite{Sanderson Ponman} are scaled using
$r_{200}$ as determined in~\cite{Sanderson 03}.  The differences
in the $r_c/r_{v,gr}$ determined using the two $M_{200}(T_X)$
relationships are within the intergroup scatter in $r_c/r_{v,gr}$.
The average profiles for the hottest and coolest groups
from~\cite{Sanderson Ponman} are well matched by using $\alpha =
6.5$ and 20 respectively and the appropriate $\beta$ and
$r_{v,gr}/r_c$. In Figures~\ref{figure 1} and~\ref{figure 2},
plots are made for stripping in $10^{13}$, $10^{14}$, and
$10^{15}M_{\odot}$ clusters. A cluster with $M_{v,gr} =
10^{15}M_{\odot}$ falls into the hottest temperature bin
from~\citet{Sanderson Ponman} while a group with
$M_{v,gr}=10^{13}M_{\odot}$ falls into the coolest. The ICM
parameters for the $10^{14}M_{\odot}$ cluster are $\alpha=5.5$,
$\beta=.57$, and $r_{v,gr}/r_c=25$.

Scatter in these parameters is estimated using the three sets of
observations.   The observed scatter in $\beta$ is $\pm0.08$. The
majority of clusters have $r_{v,gr}/r_c$ between 7 and 20, and
groups have $r_{v,gr}/r_c$ between 75 and 270.  For both groups
and clusters, $\alpha$ is assumed to vary by a factor of $\sqrt6$
in either direction. Varying $\beta$ varies the ram pressure at
$s_{orbit}=.35$ by a factor of 2 for the clusters and a factor of
7.4 for the groups. The ram pressure varies by a factor of 6 in
the clusters and 5.4 in the groups when $r_{v,gr}/r_c$ is varied.
Varying $\alpha$ by a factor of 6 results in an increase in the
ram pressure by the same factor.

\subsection{Orbits}\label{sec orbits}
The orbital parameters are important for determining both
satellites' pericenters and orbital speeds. In theory, the
pericenter of a satellite's orbit is determined by $\epsilon /
\epsilon_v$ and $b_s$ (Eq.~\ref{turningpoint}).
However,~\cite{Gill et al} present pericenter distributions for
subhalos at $z=0$ in cluster-sized dark matter halos with
$M_{v,gr}=1-3\times 10^{14}M_{\odot}$.  They find that the $s_0$
distribution can be fit with a Gaussian with $\bar{s_0}=.35$ and
$\sigma_{s_0}=.12$. This distribution shows little variation with
either $M_{v,gr}$ or $M_{v,sat}$.  The maximum ram pressure in a
cluster with ICM parameters corresponding to a few
$10^{14}M_{\odot}$ and $c_g=7.5$ increases by a factor of 5.4 when
the pericenter decreases from $s_0=0.47$ 0.23.

Some satellites have not yet passed through the pericenter of
their orbits, and the maximum ram pressure they have experienced
is determined by their current $s_{orbit}$.  Increasing
$s_{orbit}$ from 0.35 to 1 results in a decrease of the mass at
which a satellite can be stripped by two orders of magnitude.
Therefore, unless it is known that a group of satellites have all
passed through their pericenters, the effect of varying
$s_{orbit}$ cannot be treated like the scatter in the other
parameters. Instead, when comparing satellites in different
environments or with different masses, $s_{orbit}$ must be
specified.

At a given $s_{orbit}$, the ram pressure is dependent on $\epsilon
/ \epsilon_v$, but not $l/l_v$.  In simulations \citet{Vitvitska
et al.} find that the speeds of satellites entering dark halos can
be fit with a Maxwell-Boltzmann distribution with
$v^2_{rms}\approx1.15v_c^2$, $v_{c}^{2}\equiv GM_{v,gr}/r_{v,gr}$.
The total energy per unit mass of an object at the virial radius
with this speed is $\epsilon\approx\epsilon_v$, and is a weak
function of $c_g$. This distribution has a large scatter,
$\sigma_v = .6v_c$, corresponding to $\epsilon/\epsilon_v$ between
0.2 and 2.8. At $s_{orbit}=.35$, when $v_{sat}(s_{orbit}=1)/v_c$
is varied between 0.55 and 1.75, the $M_{v,sat}/M_{v,gr}$ needed
for stripping vary by a factor of $\approx 1.6$. \cite{Vitvitska
et al.} also find that for $M_{v,sat}<.4M_{v,gr}$ the angular
momenta of satellites entering halos can be fit with a
Maxwell-Boltzmann distribution with $l_{rms}/l_v=.7$, $l_v\equiv
v_cr_v$.  As expected, the distributions of $\epsilon/\epsilon_v$
and $b_s$ do not depend on $M_{v,sat}$ and $M_{v,gr}$.  However,
unless the correlation between $v_{sat}(s_{orbit}=1)/v_c$ and
$l/l_v$ is understood, these parameters cannot be used to
determine the distribution of $s_0$.

\subsection{Inclination}
\label{sec inclination}

The Gunn and Gott condition assumes that all galaxies orbit in a
face-on wind.  In reality, galaxies orbit at all inclinations.
Simulations show that galaxies are stripping at all inclinations,
but edge-on galaxies are stripped to larger radii than face-on
galaxies.

\citet{Marcolini Brighenti D'Ercole} point out that while galaxies
that are hit face-on by the ICM wind experience the most ram
pressure stripping, galaxies in an edge-on wind experience the
most viscous stripping.  They derive an expression for the ram
pressure needed to viscously strip an edge-on galaxy to radius $R$
and treat this pressure as a restoring pressure for the edge-on
case. When they compared the new condition to the Gunn and Gott
condition, they find that any ram pressure that can completely
strip a face-on galaxy will viscously strip an edge-on galaxy and
that any ram pressure that cannot strip a face-on galaxy will not
strip an edge-on galaxy. The two models differ when a galaxy is
partially stripped. In this case edge-on galaxies are still
stripped, but to larger radii than face-on galaxies.  They run
simulations at a variety of inclination angles, $i$, and find that
the final stripping radii match the predictions of the two
conditions. When galaxies at intermediate angles are partially
stripped, they are stripped to intermediate radii.

When Marcolini et al. plot the two restoring pressures for three
galaxy masses, 0.76, 7.4, and 77.2 $\times10^9M_{\odot}$, their
Figure 3, the fractional difference between the two restoring
pressures is close to constant throughout the outer disk but
increases with the mass of the galaxy.  For the largest galaxy the
restoring force in the edge-on case is $\approx4.5$ times larger.

\subsection{Concentration of the Cluster}

Small dark matter halos tend to be more concentrated than large
dark matter halos. \citet{Bullock et. al.} use an analytical model
of the evolution of dark matter halos to describe this dependency.
For the standard $\lambda$CDM model, near
$M_{v}=M{\ast}\approx1.5\times10^{13}h^{-1}M_{\odot}$ they find
\begin{equation}
c(z=0) = 9\left(\frac{M_v}{M_\ast}\right)^{-.13}
\end{equation}
The $c_g$ used for the cluster and group models are determined
using this relation. \citet{Bullock et. al.} also see a scatter
that is as large as the evolution of the concentration over the
range $0.01M_{\ast}<M_{v}<100M_{\ast}$. To explore the scatter
introduced through $c_g$, $c_g$ is varied between 5 and 16.

The cluster concentration affects both $s_0$ and the ram pressure
at a given $s_{orbit}$ (eqs.~\ref{turningpoint} and~\ref{v sat}).
When $c_g$ is increased from 5 to 16, for
$v_{sat}^2(s_{orbit}=1)=1.15v_c^2$ and $l/l_v=0.7$, the pericenter
decreases from $s_0=0.53$ to 0.5.  This is smaller than the
scatter in the pericenter. At a given $s_{orbit}$, the orbital
speed increases slightly as $c_g$ increases. This difference
increases as $s_{orbit}$ decreases. For $s_{orbit}=0.75$, the
change is not noticeable.  For $s_{orbit}=0.35$, the
$M_{v,sat}/M_{v,gr}$ at which a satellite is stripped increases by
a factor of 1.2 when $c_g$ is increased from 5 to 16.

\subsection{Concentration of the Satellite}
In the outer H \textsc{i} disk the restoring pressure of the dark
matter halo makes a non-negligible contribution to the restoring
pressure.   More concentrated satellites have higher restoring
pressures than other satellites of the same mass. Varying $c_{s}$
between 12 and 20 varies the $M_{v,sat}/M_{v,gr}$ at which
stripping occurs by a factor of $\approx 1.7$.

The concentration also alters how the hot galactic halo is
stripped. In a more concentrated dark matter halo the gas is
denser in the center and more diffuse elsewhere.  Stripping can
therefore occur down to a smaller radius, but a smaller fraction
of the gas is lost.

\subsection{Length Ratios, $\lambda_i$}

How disk scale lengths vary with $M_{v,sat}$ is an open question.
Theory suggests that when cold gas disks form in dark matter
halos, their initial radii are proportional to the virial radii of
the halos they form in~\citep{Mo Mao White}. There is a large
scatter in this relationship that is due to a large scatter in the
disk angular momenta. However, cold disks are composed of a
combination of stars and gas, and how these components arise from
the original cold disk is not understood.

The fraction of the H \textsc{i} that is found in the outer disk,
where the restoring force from the stellar disk is weakest, is set
by the ratio $\lambda_d / \lambda_g$ (eqs.~\ref{stellar disk 2}
and~\ref{sigma not}). \citet{Swaters et. al.} observed the H
\textsc{i} disks of 73 local dwarfs.  They found $\lambda_d /
\lambda_g = 6 \pm 2.5$ for the dwarfs.  For late-type spirals
$\lambda_d / \lambda_g = 5.5 \pm 1.6$.   These observations
suggest that $\lambda_d / \lambda_g$ contributes to the scatter in
$M_{v,sat}/M_{v,gr}$, but not to the mass dependence.  Varying
$\lambda_d / \lambda_g$ changes both the size and the shape of the
restoring force.  When $\lambda_g$ is held at 5.7 and
$\lambda_d/\lambda_g$ is decreased to 3.5, the restoring force
decreases within $R/R_g = 0.45$ and increases at larger $R$. The
fractional change in $P_{rest}$ is less than 1.1 throughout the
disk. Holding $\lambda_g$ constant and increasing
$\lambda_d/\lambda_g$ to 8.5, results in decreasing the restoring
force outside of $R/R_g=0.3$ by a factor of less than 1.13.

Varying $\lambda_g$ while holding $\lambda_d / \lambda_g$ constant
changes the density of both the gas and stellar disks.  At a given
$R/R_g$, the density of both disks $\propto\lambda_g^2$ and the
restoring pressure $\propto\lambda_g^4$.  The scatter in
$\lambda_g$ and how $\lambda_g$ evolves with mass are not known.
However, because of the heavy dependence of the restoring pressure
on it, the scatter in $\lambda_g$ is likely to make an important
contribution to the scatter in the $M_{v,sat}/M_{v,gr}$ at which
stripping occurs. Increasing $\lambda_g$ from $0.75\lambda_g$ to
$1.25\lambda_g$ decreases the $M_{v,sat}/M_{v,gr}$ needed for
stripping by a factor of 20.

The presence of molecular gas complicates the physics of stripping
in the inner disk.  Therefore, this model is only applied to disk
radii beyond 1.5 stellar disk scale lengths.  At these radii,
$R>10r_b$, the bulge acts as a point mass, and changing
$\lambda_b$ (Eq.~\ref{bulge 2}) has no effect. In particular, the
radius of the bulge changes with the mass of the bulge.  Below,
the mass of the bulge is allowed to vary by 40\%.  Assuming that
$r_b\propto M_b^{1/3}$, this corresponds to allowing $r_b$ to vary
by 12\%. This is not noticeable in the outer H \textsc{i} disk.

\subsection{Mass Fractions, $m_i$}\label{mi}

Both $m_b$ and $m_{dg}$ (eqs.~\ref{bulge 2} and~\ref{sigma not})
are correlated with the mass of the satellite. While the majority
of large spiral galaxies have a significant fraction of their
stars in a stellar bulge, low-mass late-type galaxies do not. Very
low mass spiral galaxies exist that do not contain any
bulge~\citep{Matthews Gallagher} and dwarf galaxies are observed
to be either dE/dSph or dIrr. In addition, low-mass late-type
galaxies have a higher fraction of their mass in the gas disk than
large spirals~\citep{Swaters et. al.}. Decreasing $m_b$ reduces
the restoring pressure in the outer disk. In contrast, increasing
$m_{dg}$ increases the restoring pressure.

In the bottom right panels of Figures \ref{figure 1} and
\ref{figure 2}, the satellite is bulgeless, the gas mass fraction,
$m_{dg}$, is doubled, and the stellar mass fraction, $m_{ds}$, is
reduced by 20\%. The original model corresponds to a large
late-type spiral, while the bulgeless model is closer to a
late-type dwarf galaxy. Because the restoring force due to the
bulge is small, the net effect is to increase the restoring
pressure. Most galaxies will lie between the two plotted models.

The $m_i$ can be split between those that set the depth of the
potential well, $m_{ds}+m_{H_2}$ and $m_b$, and $m_{dg}$ which
sets the gas density.  In late-type galaxies, bulges can
contribute up to 50\% of the stellar light~\citep{Binney
Merrifield}. In the reference model, 25\% of the stars are in the
bulge. Varying $m_b$ between 15\% and 35\% of the satellite's
stars, varies the restoring force in the outer disk by a factor of
$\approx1.1$. Neither the mass dependence nor the scatter in
$m_{ds}+m_{H_2}$ are known. Varying $m_{ds}+m_{H_2}$ by 20\%,
varies the restoring force by a factor of $\approx 1.3$. The
effect of scatter in the gas density parameter, $m_{dg}$, is
straight forward.  The $M_{v,sat}/M_{v,gr}$ at which stripping
occurs varies as $m_{dg}^{-3/2}$. Varying $m_{dg}$ by a factor of
1.5, similar to the scatter in $M_{H \textsc{i}}/L_R$ in Figure 9
of Swaters et. al., varies the $M_{v,sat}/M_{v,gr}$ needed for
stripping by a factor of 1.8.

Variations in the $m_i$ values should be correlated because the
variation in the mass in baryons in a satellite is likely smaller
than the variation in the mass of a given component.  The
correlation is such that it will reduce the overall scatter.

\section{DISCUSSION}

\begin{figure}
\includegraphics[scale = 0.80]{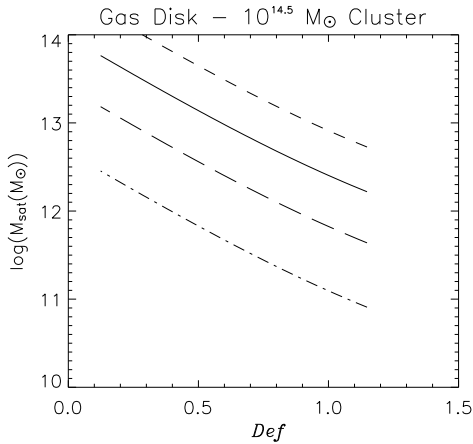}\caption{Mass at which a
satellite's gas disk is ram pressure stripped versus the
deficiency in a $3.5\times10^{14}M_{\odot}$ cluster. Lines are the
same as in Fig. 1.}\label{figure 4}
\end{figure}

\subsection{Comparison with Observations and Simulations}
In this section, the $M_{v,sat}/M_{v,gr}$ values at which the
model predicts stripping should occur are compared to observations
of H \textsc{i} disks from~\cite{Solanes et al 2001}.  This sample
is useful for making comparisons between the model and
observations because the number of spirals is large and the
observations span several clusters. The average temperature of the
clusters is 3.5keV. This corresponds to
$M_{v,gr}=3\times10^{14}M_{\odot}$. The average gas profile for
this temperature can be described using the middle mass cluster
model. In Figure 4 from Solanes et al. the fraction of spirals
with $def>0.3$ and the average deficiency are plotted versus
$r/R_A$, where $R_A$ is the Abel radius. In Figure~\ref{figure 4}
of the present paper $M_{v,sat}$ is plotted versus the deficiency
in a $3\times10^{14}M_{\odot}$ cluster at several $s_{orbit}$. The
deficiency is defined as $def\equiv\log(\bar{M}/M_{obs})$, where
$M_{obs}$ is the observed H \textsc{i} mass and $\bar{M}$ is the
average H \textsc{i} mass for field spirals with the same
morphology and optical diameter. For the model
$def=log(M_g/(M_g-M_{str}))$.

The observed fraction of spirals with $def>0.3$ reaches 0.3 at
$r/R_A\approx1.$ and 0.45 at $r/R_A\approx0.5$.  The model
predicts that almost all galaxies within $s_{orbit}=0.5$ have
$def>0.3$. However, only $\approx30\%$ of the galaxies viewed
within $r/r_{v,gr}=0.5$ actually reside within $r/r_{v,gr}=0.5$.
This estimate assumes that the galaxy distribution follows an NFW
profile and that all galaxies within $r/r_{v,gr}=2$ are included
in the projected cluster.  At $s_{orbit}=1$,  the model predicts
that an average galaxy traveling face-on to the wind is stripped
to $def>0.3$ for $M_{v,sat}<10^{11.5}M_{\odot}$. This estimate can
be roughly adjusted for inclination by multiplying by $10^{-.5}$,
$M_{v,sat}<10^{11}M_{\odot}$.  Approximately 45\% of the galaxies
seen within $s_{orbit}=1$ actually reside within $s_{orbit}=1$.
Some galaxies beyond the $s_{orbit}$ under consideration will be
stripped to $def>0.3$.  Therefore, the observed fraction of
spirals with $def>0.3$ should be greater than that predicted by
multiplying the volume fraction with the fraction of stripped
galaxies within $s_{orbit}$. This condition is met for
$s_{orbit}=1$ if at least a third of the observed spirals have
$M_{v,sat}>10^{11}M_{\odot}$, corresponding to
$M_{d}\approx8\times10^{9}M_{\odot}$. For $s_{orbit}=0.5$, the
volume fraction is lower than the observed fraction and this
condition is easily met. The difference between the observed value
and product of the volume fraction and stripped fraction should
decrease as $s_{orbit}$ increases.  A $10^{11}M_{\odot}$ spiral
traveling face-on to the wind is stripped to $def\approx1$ at
$s_{orbit}=0.8$. A $10^{12}M_{\odot}$ spiral traveling face-on to
the wind is stripped to $def\approx1$ at $s_{orbit}=0.4$.  Spirals
with $def\approx1$ are not observed in the~\citet{Solanes et al
2001} sample outside of $1R_A$.

The model does a good job of matching the observations of H
\textsc{i} disks from~\citet{Solanes et al 2001}.  If the model
over-predicted the degree of stripping, then it would over-predict
the fraction of spirals with $def>0.3$ and predict deficiencies at
large radii greater than those observed. If the model
under-predicted the degree of stripping, then it might not predict
that spirals in the center of the cluster can have $def\approx1$
and would require that a larger fraction of the observed spirals
reside in the cluster center.

Both~\citet{Marcolini Brighenti D'Ercole} and~\citet{Roediger and
Hensler} run simulations of disk galaxies in winds typical of the
outskirts of a group, equivalent to $s_{orbit}=1$ and 0.7, and the
center of a group, $s_{orbit}=0.2$ and 0.25. In Marcolini et al.
the two lowest mass simulated dwarfs are completely stripped in
the high ram pressure. In the other cases, the dwarfs are
partially stripped to a range of radii.  When the gas surface
density in the current model is adjusted to the exponential gas
density in the dwarf models, the current model and the simulations
match well for both $s_{orbit}$. The only disparity is that for
$s_{orbit}=0.2$ the high-mass model is stripped to $R_{str}/R_{g}
= 0.1$ in the simulations and to $R_{str}/R_{g}=0.45$ in the
model. This is probably due to differing dark matter halo
profiles.  The high-mass, $M>10^{11}M_{\odot}$, galaxy model of
Roedinger and Hensler is stripped to $R_{str}/R_{g}=0.8$ in the
lower ram pressure and $R_{str}/R_{g}=0.6$ in the higher ram
pressure. After the gas density is adjusted, the stripping radii
of their high-mass model in the two winds are matched well by the
model.

\subsection{Model Predictions}
\label{predictions}

In this model there exist three stripping regimes, stripping of
the hot galactic halo, the flat H \textsc{i} disk, and the inner
gas disk. The focus is on stripping of the flat H \textsc{i} disk.
However, the model also includes a hot galactic halo.  This hot
gas is easily stripped in all environments. A satellite galaxy
that is orbiting at $s_{orbit}=1.$ and has a hot galactic halo
consisting of 0.5\% of its mass, is stripped of at least $40\%$ of
its hot halo.  At $s_{orbit}=0.35$, a typical pericenter, all
satellite galaxies are stripped of practically their entire halo
in both groups and clusters. The stripping of the inner disk has
not been studied. However, this gas should be difficult to strip
and any spiral that has not been stripped of its entire outer H
\textsc{i} disk should retain this gas.  The rest of this section
focuses on the outer disk.

The model does a reasonable job of matching H \textsc{i}
observations in large cluster spirals. Therefore, it can be used
to predict the extent of stripping at different $M_{v,sat}$ and
$M_{v,gr}$.  The model's basic prediction, that the stripping
radius is determined by the ratio $M_{v,sat}/M_{v,gr}$ is modified
by the mass dependence of the model's parameters. Therefore, the
usefulness of the model's predictions depends in part on how well
this dependence is understood.  Fortunately, the mass dependence
of many of the parameters is based on observations. The model
parameters that most alter $M_{str}(M_{v,sat}/M_{v,gr})$ in
low-mass systems are the ICM profile parameters and the
distribution of mass in the galaxy. The scaled ram pressure,
$P_{ram}/M_{v,gr}^{2/3}$, that a satellite experiences decreases
with $M_{v,gr}$ because the ICM density throughout most of the
cluster decreases. The scaled restoring pressure,
$P_{rest}/M_{v,sat}^{2/3}$, is larger in low-mass disks because a
higher fraction of the disk mass is found in the H \textsc{i}
disk. The changes in both the scaled ram and restoring pressures
act to decrease the $M_{v,sat}/M_{v,gr}$ at which stripping occurs
in low-mass systems. For example, the Local Group dwarfs have
$M_{v,sat}/M_{v,gr}$ smaller than the bright spirals in Virgo but
should experience similar amounts of stripping.

The model makes three basic predictions.  Dwarf galaxies are
stripped of their entire outer H \textsc{i} disks in clusters and
lose varying fractions of their outer disk in groups.  Massive
spirals can also be stripped of significant fractions of their H
\textsc{i} disk in groups if they travel to small $s_{orbit}$. In
a $10^{13}M_{\odot}$ group, a $10^{10}M_{\odot}$ dwarf is stripped
of 50\% of its H \textsc{i} disk at $s_{orbit}=0.7$, a
$10^{11}M_{\odot}$ satellite is similarly stripped at either
$s_{orbit}=0.5$ or $s_{orbit}=0.35$, depending on the satellite
model used, and a large $5\times10^{11}M_{\odot}$ satellite is
similarly stripped at $s_{orbit}=0.3$.  The corresponding orbits
from stripping 80\% of the H \textsc{i} disk are $s_{orbit}=0.55$,
0.35, 0.25, and 0.2.  It should be kept in mind that these numbers
are for an average satellite traveling face-on to the ICM wind.

The degree of stripping that a particular satellite of mass
$M_{v,sat}$ experiences orbiting in a cluster of mass $M_{v,gr}$
can vary greatly.  The parameters that are responsible for most of
the variation in $M_{str}$ are the ICM parameters, $\alpha$,
$\beta$, $r_{v,gr}/r_c$, the galaxy's inclination, $i$, the
pericenter, $s_0$, and the extent of the disk, $\lambda_g$. The
variation in the stellar and gas surface densities is largely
contained in $\lambda_g$.  The ICM parameters are set by the
choice of cluster, but clusters contain galaxies with a variety of
orbits, morphologies, and inclinations. Within the same cluster, a
galaxy of mass $M_{v,sat}=M$ with a low surface density on a
radial orbit traveling face-on to the wind can be as severely
stripped as a galaxy with $M_{v,sat} \approx M/25$ that has a high
surface density and is traveling edge-on.  At the same time, the
ram pressure can vary significantly between clusters with the same
mass because of variations in the ICM profile.  Two identical
satellite galaxies on identical orbits in different clusters of
the same mass can experience ram pressures that vary by a factor
of $\approx10$.

An effective mass, $M_{eff}$, can be defined for each individual
galaxy.  This is the mass of a satellite that the model predicts
is stripped to the same $R_{str}$ as the given galaxy.  The effect
of the scatter in the parameters can then be discussed in terms of
the scatter in $M_{eff}$ at a given $M_{v,sat}$.  The scatter in
the $M_{v,sat}/M_{v,gr}$ at which a galaxy is stripped of
$M_{str}$, or in $M_{eff}$, can best be expressed as a
multiplicative factor,
$log(M_{eff})=log(M_{v,sat}/M_{v,gr})\pm\sigma_{log}$. The source
of scatter in $M_{eff}$ can be grouped into three types, that due
to the ICM parameters, $\alpha$, $\beta$, $r_{v,gr}/r_c$, the
orbital parameters, $\epsilon/\epsilon_v$, $i$, $c_g$, and the
restoring pressure parameters, $c_s$, $m_i$, $\lambda_i$.  The
overall scatter in the high- and low-mass models, the result of
varying all of the model parameters, are
$\sigma_{log,h}\approx0.75$ and $\sigma_{log,l}\approx0.85$. These
can be broken down into $\sigma_{log,ICM,h}=0.7$,
$\sigma_{log,ICM,l}=0.8$, $\sigma_{log,orbit}=0.5$, and
$\sigma_{log,rest}=0.65$.  The discussion in the paragraph above
is based on these estimates.

The uses of this model are twofold. First, it makes general
predictions about the degree of stripping that occurs to galaxies
of different masses in a range of environments.   Second, it links
the degree of stripping, $M_{str}$ or $R_{str}$, to the ratio
$M_{v,sat}/M_{v,gr}$.  This may be useful for verifying whether
stripping is occurring within a galaxy catalog.  The size of the
scatter in $M_{eff}$ determines the size of the galaxy catalog
that is needed to see a dependence of $M_{str}$ on
$M_{v,sat}/M_{v,gr}$.  The scatter in the $\beta$ profile
parameters is a major contributor to the scatter in $M_{eff}$.
Therefore, such a catalog must either contain many clusters, or
the ICM profile of the clusters must be known.  As an estimate of
the catalog size necessary, if 200 satellite galaxies with
approximately the same $M_{v,sat}$ orbiting in 50 groups or
clusters with the same $M_{v,gr}$ are combined, then
$log(\bar{M}_{eff})\approx log(M_{v,sat})\pm.25$.  If the range in
$M_{v,sat}$ and $M_{v,gr}$ of such a catalog is large compared to
this scatter, then the effects of stripping should be seen in the
relationship between $M_{str}$ and $M_{v,sat}/M_{v,gr}$.

\section{CONCLUSIONS}

The model developed here relates the degree of ram pressure
stripping a satellite galaxy experiences to the galaxy and cluster
masses and can be used to quickly determine the extent to which a
galaxy is likely to be stripped.  All galaxies lose most of their
hot galactic halo. In clusters galaxies at moderate mass ratios,
$M_{v,sat}/M_{v,gr}\approx10^{-2}$, are moderately stripped,
$R_{str}/R_d\approx0.6$, at intermediate distances, $0.5 <
s_{orbit} < 1$, from the cluster center and severely stripped
closer to the cluster center. Satellites with lower
$M_{v,sat}/M_{v,gr}$ are severely stripped even at intermediate
distances. Stripping also occurs in groups.  However, the same
degree of stripping occurs for lower $M_{v,sat}/M_{v,gr}$ in
groups than in clusters.  Dwarf galaxies are moderately to
severely stripped at intermediate distances in groups, and large
spiral galaxies can be moderately stripped if they travel to small
$s_{orbit}$.

The model is simple and motivated by observations. Dark matter
profiles are well matched to NFW profiles outside a possible core,
observations of X-ray gas in clusters can be fit using $\beta$
profiles, and the average gravitational potential of most galaxies
should be well matched by the model potential.  The model has a
large number of parameters, but the values for many can be taken
from observations.  However, the model assumes that clusters are
static.  From an evolutionary standpoint, it is only valid after
the group or cluster has acquired an ICM. On the other hand, in
dynamic clusters bulk ICM motions may cause more stripping to
occur than is predicted by the model. This is observed in the
Virgo cluster~\citep{Kenney van Gorkom Vollmer 2004}.

Ram pressure stripping is not the only way to remove gas from
galaxies.  In groups and clusters tidal stripping of both gas and
stars can occur (e.g.,~\citep{Bureau et al, Patterson Thuan}). For
dwarf galaxies it has been proposed that supernova winds
associated with bursts of star formation may expel
gas~\citep{Ferrara Tolstoy, Silich Tenorio-Tagle}, and the
efficiency of this mechanism may depend on the
environment~\citep{Murakami Babul}.  However, ram pressure
stripping is capable of removing gas from galaxies across a large
range of galaxy masses and environments. In particular ram
pressure stripping can act when supernova-driven ejection is
likely to be inefficient and can remove gas from galaxies that are
either not tidally interacting or that are experiencing only weak
tidal interactions.

The tidal radius, $r_t$, at the physical orbital radius,
$r_{orbit}$, as estimated using $r_t =
(M_{v,sat}/3M_{enc})r_{orbit}$, can be compared to the stripping
radius, $R_{str}$, as determined using the model.  Here $M_{enc}$
is the cluster mass enclosed within $r_{orbit}$.  In scaled
coordinates, $r_t/R_d=\lambda_g s_{orbit}[3m(s_{orbit})]^{-1/3}$,
where $m(s_{orbit})=M_{enc}/M_{v,gr}$.  The H \textsc{i} fraction
lost to tidal stripping depends only on the scaled size of the gas
disk and $s_{orbit}$ and not on either mass. Because the gas disk
resides in the center of the satellite's dark halo, tidal
stripping rarely effects the H \textsc{i} disk and practically
never competes with the effect RPS.  For example, in order for the
H \textsc{i} disk to be tidally stripped to $r_t/R_g =.8$, a
satellite must travel to $s_{orbit}\approx.1$. By this $s_{orbit}$
the entire outer H \textsc{i} disk has been ram pressure stripped
for almost all satellites.

The effect of varying the model's parameters was studied both to
identify the parameters that most effect the extent to which an
individual galaxy is stripped and to determine the range of
satellite and cluster masses that result in the same $R_{str}$.
The extent to which an individual galaxy is stripped depends most
strongly on the galaxy's $s_{orbit}$ or $s_0$, inclination, $i$,
and disk scale length, $\lambda_g$, and on the cluster's ICM
profile.  Galaxies that reach smaller $s_{orbit}$, are face-on to
the wind, have denser disks, or orbit through a denser ICM are
more effectively stripped.  Galaxies in the same cluster with
$M_{v,sat}$ that differ by as much as a factor of 25 can be
stripped to the same $R_{str}$.  In different clusters, galaxies
on identical orbits with identical morphologies, identical $m_i$,
$\lambda_i$, $c_s$, can be stripped to the same $R_{str}$ with
$M_{v,sat}$ values that differ by as much as a factor of 30.

Ram pressure stripping of satellite galaxies' outer gas disks and
hot galactic halos is occurring frequently in both groups and
clusters. In general, removing gas from a galaxy reduces star
formation. The gas in the outer disk and hot galactic halo is not
currently involved in star formation, but may feed future star
formation if it is not stripped. Therefore, it would be
interesting to study in more detail how the removal of this gas
affects galaxy evolution and to determine if, to what extent, and
how quickly star formation declines after a galaxy is stripped.

\section{ACKNOWLEDGEMENTS}
This project was advised by D. N. Spergel and funded by NASA Grant
Award \#NNG04GK55G.   I'd like to thank the anonymous referee for
useful comments.

\appendix

\section{NFW Profile}
\label{NFW profile}

When simulations are run that allow cold, non-interacting
particles to evolve gravitationally from a flat power spectrum at
high redshift, the dark matter halos that emerge at low $z$ are
well fit by a universal density profile over many orders of
magnitude in mass. This profile was originally parameterized by
\citet{NFW}. Both the satellite and group/cluster dark matter
halos are described using an NFW profile, the relevant properties
of which are summarized here.

The NFW profile is
\begin{equation}
\frac{\rho (r)}{\rho ^{0}_{c}}=\frac{\delta
_{char}}{(r/r_{s})(1+r/r_{s})^{2}}
\end{equation}
where $\rho_c^0$ is the average density of the universe at $z=0$,
$r_s$ is an inner scale radius, and $\delta_{char}$ is a
characteristic over-density.  It can be rewritten in terms of a
concentration, $c$, an overdensity, $v_{\rho}$, and a scaled
radius, $s\equiv r/r_v$
\begin{equation} \label{NFW}
\frac{\rho(s)}{\rho_{c}^{0}}=\frac{v_{\rho}c^2g(c)}{3s(1+cs)^{2}}
\end{equation}
\begin{equation}
c\equiv \frac{r_{v}}{r_{s}}
\end{equation}
\begin{equation}\label{delta}
\delta_{char}=\frac{v_{\rho}c^{3}g(c)}{3}
\end{equation}
\begin{equation}
g(c)=\frac{1}{\ln(1+c)-c/(1+c)}
\end{equation}
The overdensity, virial mass, $M_v$, and virial radius, $r_v$, are
related by
\begin{equation}\label{Mv of rv}
M_{v}=\frac{4}{3}\pi r^{3}_{v}v_{\rho}\rho ^{0}_{c}
\end{equation}
Note that the virial radius scales as $M_{v}^{1/3}$. The mass
contained within $s$ is
\begin{equation}
M(s)=g(c)M_{v}\left[\ln(1+cs)-\frac{cs}{1+cs}\right]
\label{NFWMass}
\end{equation}
and the gravitational potential is
\begin{equation}
\Phi (s) = -\frac{GM_{v}g(c)}{r_v}\frac{\ln(1+cs)}{s}
\end{equation}
The above can be found in \citep{Cole Lacey, Lokas Mamon}
and~\citep{NFW}.

\section{Orbits in an NFW potential}
\label{ap orbits} Bound orbits in a central force field travel
between two radial extremes. These are found by solving.
\begin{equation}
\frac{1}{r^{2}}=\frac{2\left[\epsilon - \Phi (r)\right ]}{l^{2}}
\end{equation}
where $\epsilon$ and $l$ are the energy and angular momentum per
unit mass~\citep{Binney Tremaine}.

The scaled radial extremes, $s_0$, for an NFW profile are given by
\begin{equation}
\frac{1}{s_{0}^{2}}=\frac{-2}{b_{s}^{2}}\left[1-\frac{\epsilon_{v}}{\epsilon}g(c)
\frac{\ln(1+cs_{0})}{s_{0}}\right] \label{turningpoint}
\end{equation}
where $\epsilon_v$ and $b_s$ are defined as
\begin{equation}\label{orbital energy}
\epsilon _{v} \equiv -\frac{GM_{v}}{r_{v}}
\end{equation}
\begin{equation}\label{impact parameter}
b_{s}^{2}\equiv
\frac{l^{2}}{|\epsilon|r^{2}_{v}}=\frac{b^{2}}{r^{2}_{v}}
\end{equation}
The parameter $b$ would be the impact parameter for an unbound
orbit.

Conserving angular momentum and using Eq. \ref{turningpoint}, the
orbital speed at the pericenter, $v_0$, is
\begin{equation}\label{max velocity}
v_{0}^{2}=\frac{b^{2}_{s}}{s_{0}^{2}}\frac{\epsilon}{\epsilon_{v}}\frac{GM_{v}}{r_{v}}
\end{equation}
It is proportional to $M_{v}^{1/3}$ and depends on the
concentration through $s_{0}(c)$.

At any point in the orbit, the orbital speed, $v_{sat}$, is
\begin{equation}\label{v sat}
v_{sat}^2=\frac{2GM_{v}}{r_{v}}\left[g(c)\frac{\ln(1+cs)}{s}-\frac{\epsilon}{\epsilon_{v}}\right]
\end{equation}

\section{Hot Galactic Halo}
\label{ap satellite halo}

The virial temperature, $T_v$, of a gravitational potential is
defined by
\begin{equation}
\label{virial temp} \frac{3k_{B}T_v}{\mu m_{p}}\equiv\langle
v^{2}\rangle =\frac{|W|}{M_v}
\end{equation}
where $W$ is the work done by gravity in forming the halo, $k_B$
is the Boltzmann constant, $m_p$ is the mass of the proton, and
$\mu$ is the mean molecular weight~\citep{Binney Tremaine}.

Using eqs. \ref{NFW}, \ref{Mv of rv}, and \ref{NFWMass}, the
virial temperature of an NFW profile is
\begin{equation}
\frac{|W|}{M_v}=\frac{GM_{v}}{r_{v}}c^{2}g^2(c)
\int^{1}_{0}\frac{\ln(1+cs)-cs/(1+cs)}{s(1+cs)^{2}}sds
\end{equation}
\begin{equation}
\label{work}
\equiv \frac{GM_{v}}{r_{v}}c^{2}g(c)I_{1}(c)
\end{equation}
A dimensionless temperature, $t$, is defined.
\begin{equation}
t(c)\equiv \left(\frac{k_B T_v}{\mu m_p}\right) \left(\frac{r_v}{G
M_v}\right)
\end{equation}
\begin{equation}
t(c)=\left(\frac{c^2g(c)I_1(c)}{3}\right)
\end{equation}

For a gas in hydrostatic equilibrium in potential $\phi$, ignoring
self-gravity
\begin{equation}
\label{hydro eq}
 \rho(r)=\rho_{0}\exp\left({\frac{\mu
 m_{p}}{k_{B}T}}\left[\phi (r) - \phi (0) \right]\right)
\end{equation}

Combining eqs. \ref{work} and \ref{hydro eq}, the gas density
profile is
\begin{equation}
\rho_{g}(s)=\rho_{0}\exp\left(\frac{3}{cI_{1}(c)}\left[\frac{\ln(1+cs)}{cs}-1
\right]\right)
\end{equation}
Letting the gas mass within $r_{v}$ equal a fraction $m_{sg}$ of
the virial mass results in a central density of
\begin{equation}
\rho_{0}=\frac{\rho_{0}^{c} v_{\rho} m_{sg}}{3I_{2}(c)}
\end{equation}
The integral, $I_{2}(c)$, is
\begin{equation}
I_{2}(c)
=\int_{\epsilon}^{1}s^{2}ds\exp\left(\frac{3}{cI_{1}(c)}\left[\frac{\ln(1+cs)}{cs}-1\right]\right)
\end{equation}

The scaled density profile $j$ is defined by
\begin{equation}
j(s,c)\equiv\frac{1}{I_2(c)}\exp\left(\frac{3}{cI_1(c)}\left[\frac{\ln(cs)}{cs}-1\right]\right)
\end{equation}

\begin{equation}
\rho(s) = \left(\frac{v_{\rho}\rho_0^c}{3}\right)m_{sg}j(s, c)
\end{equation}

\begin{deluxetable}{clccclcc}
\tabletypesize{\scriptsize} \tablecaption{Parameter Values and
Scatter} \tablewidth{0pt} \tablehead{ \colhead{parameter} &
\colhead{ref in text} & \colhead{reference model} &  \colhead{mid
mass model} & \colhead{low mass model} & \colhead{scatter} &
\colhead{$\Delta P$}\tablenotemark{a} & \colhead{$\Delta
M$}\tablenotemark{a} } \startdata

$\alpha$                    & Eq.~\ref{beta}   & 6.5   & 5.5   & 20    & $\times6$   & 6      & 15   \\
$\beta$\tablenotemark{b}    & Eq.~\ref{beta}   & 0.67  & --    & --    & $\pm.08$    & 2      & 3    \\
--                          & --               & --    & 0.57  & --    & --          & --     & --   \\
--                          & --               & --    & --    & 0.45  & $\pm.08$    & 7.4    & 20   \\
$r_{v,gr}/r_c$\tablenotemark{b} & Eq.~\ref{beta} &12   & --    & --    & 7 - 20      & 6      & 16   \\
--                          & --               & --    & 25    & --    & --          & --     & --   \\
--                          & --               & --    & --    & 170   & 75 - 270    & 5.4    & 13   \\
$i$ & sec.~\ref{sec inclination} & $0^{\circ}$ & -- & -- &$0^{\circ}$ - $90^{\circ}$ & 4.5    & 10   \\
$v_{sat}(s_{orbit}=1)/v_c$  & Eq.~\ref{v sat}  & 1.15  & --    & --    & $\pm.6$     & 1.4    & 1.6  \\
$c_g$                 & Eq.~\ref{group NFW}    & 5.5   & 7.5   & 10    & 5-16        & 1.1    & 1.2  \\
$c_s$ & Eq.~\ref{eqn. full restoring pressure} & 12    & --    & 20    & 12-20       & 1.4    & 1.7  \\
$\lambda_g$              & Eq.~\ref{sigma not} & 5.7   & --    & --    & $\pm1.4$    & 8      & 20   \\
$\lambda_d/\lambda_g$ & Eq.~\ref{stellar disk 2} & 6   & --    & --    & $\pm2.5$    & 1.2    & 1.4  \\
$\lambda_h$         & Eq.~\ref{stellar disk 2} & 0.04  & --    & --    & --          & --     & --   \\
$\lambda_b$                & Eq.~\ref{bulge 2} & 180   & --    & na    & --          & --     & --   \\
$m_{dg}$                 & Eq.~\ref{sigma not} & 0.008 & --    & 0.016 &$\times$1.5  & 1.5    & 1.8  \\
$m_{ds}+m_{H_2}$    & Eq.~\ref{stellar disk 2} & 0.05  & --    & 0.04  &$\pm .01$    & 1.3    & 1.4  \\
$m_b$                      & Eq.~\ref{bulge 2} & 0.017 & --    & 0     &0.012-0.024  & 1.1    & 1.2  \\
$m_{sg}$                &Eq.~\ref{g halo prof} & 0.005 & --    & --    &$\times$1.5  & 1.5    & 1.8  \\
$s_{0}$                &Eq.~\ref{turningpoint} & 0.35  & --    & --    &$\pm.12$     & 5.4    & 13   \\
\enddata

\tablenotetext{a}{$\Delta P$ and $\Delta M$ columns list the
scatter in $P_{ram}$ or $P_{rest}$ and $M_{v,sat}/M_{v,gr}$.  The
scatter is given as the fractional change when the parameter of
interest is varied over its entire range.  Therefore, $\log(\Delta
M) \approx 2\sigma_{log}(M_{v,sat}/M_{v,gr})$.}

\tablenotetext{b}{The first row shows affects of scatter for the
large mass model and the third row for the small mass model.
Columns 7 and 8 show the fractional increase in $P_{ram}$ at
$s_{orbit}=1$.}
\end{deluxetable}

\end{document}